# Period-doubling route from synchronization to chaos of an oscillator coupled to a regular oscillator

Igal Berenstein*

Tr 18A # 96-10 Ap 501, Bogotá, Colombia

**Abstract**

Spatiotemporal chaos in the form of defect-mediated turbulence is known for oscillators coupled by diffusion. Here we explore the same conditions that produce defect turbulence, in an array of oscillators that are coupled through the activator to a regular oscillator. We find that for very small coupling the oscillators behave independent of each other and then there is a transition to complete synchronization. On further increasing the coupling strength, there is period doubling and a transition to chaotic behavior of each driven unit. However the global behavior shows some ordering, and the period-two oscillations become period-one with a further increase in the coupling strength.

* e-mail: igalb@excite.com

# I. Introduction

Coupling of oscillators is a subject of intense research and for a recent review see the paper of Pikovsky and Rosenblum [1], and in chemistry has received some attention recently. The most common of chemical oscillators is the Belousov-Zhabotinsky chemical oscillator in it has been used in studies of coupled oscillators [2, 3] but different types of oscillators have also been used as well, such as the mercury beating heart [4]. The results of such studies can be applied to building chemical computers [5].

The expected main result is that for coupled oscillators, whether identical or not, when the coupling is strong enough, the oscillators will synchronize [1, 6], which is also seen in systems with delay [7]. However, reaction-diffusion systems, that can be seen as oscillators coupled to its nearest neighbors *via* diffusion, can produce spatiotemporal chaos. This is exemplified by amplitude equations close to a bifurcation point that take the form of the Complex Ginzburg-Landau equation [8]. It is also seen in common models for oscillating chemical reactions such as the *Oregonator* model [9] and the Gray-Scott model [10] within the oscillatory domain. The form of spatiotemporal chaos observed can take the form of *defect-mediated turbulence,* in which the amplitude of the patterns vanishes at some points in space and time. The transition from bulk (synchronized) oscillations to defect-mediated turbulence occurs through a type of chaos known as phase chaos where there are no breaks in phase [8]. Is this transition the only type of transition that can happen? Here we explore the behavior of an array of oscillators with *Oregonator* kinetics coupled to a single unperturbed oscillator. We use this model that comes from the kinetics of the Belouzov-Zhabotinsky reaction [11], and since this reaction is known to produce *defect-mediated turbulence* [12], the results of our simulations can be compared to experiments. Section II describes the model we use, section III shows the results and the transition to chaos, section IV is devoted to discussion and section V to conclusions.

# II. The model

The model we use here reads:

$$\partial x / \partial t = (1/\varepsilon)(x(1-x) - fz(x-q)/(x+q)) + C(x_b - x)$$
$$\partial z / \partial t = (x - z)$$
$$\partial x_b = (1/\varepsilon)(x_b(1-x_b) - fz_b(x_b - q)/(x_b + q))$$
$$\partial z_b / \partial t = (x_b - z_b)$$

where $x$ represents the concentration of the activator (HBrO$_2$) for each oscillator, and $z$ the catalyst (ferroin) which acts as an inhibitor and is immobilized within each oscillator. The activator can diffuse to a common bath for all oscillators as seen in experiments [13,14]. $C$ corresponds to a coupling strength while $x_b$ corresponds to the concentration of the activator in the common bath. The parameters are chosen as $f = 2.1$, $q = 0.017$ and $\varepsilon = 0.04$, which correspond to defect-mediated turbulence for a reaction-diffusion system [9]. The system is integrated using the Euler method with a time step $\Delta t = 0.001$ time units (t. u.). The common bath is assumed to be big in comparison to the oscillators, so that the exchange with the oscillators does not change the concentration of the common bath. This is illustrated in figure 1.

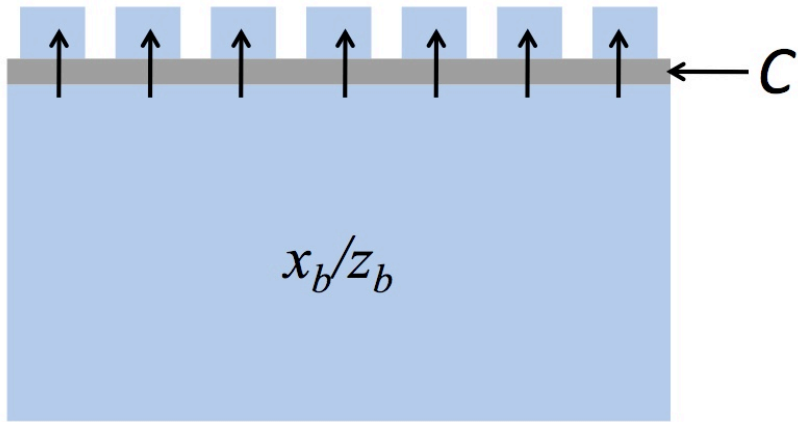

**Figure 1: Schematic of the simulation.**

The coupling strength corresponds to changing the thickness of the boundary between the common bath and the oscillators, as has been done in coupled layers displaying Turing patterns [15].

## III. Results

Figure 2 shows a complete phase diagram of the patterns obtained for a network consisting of 500 units plus the common bath by changing the coupling strength $C$.

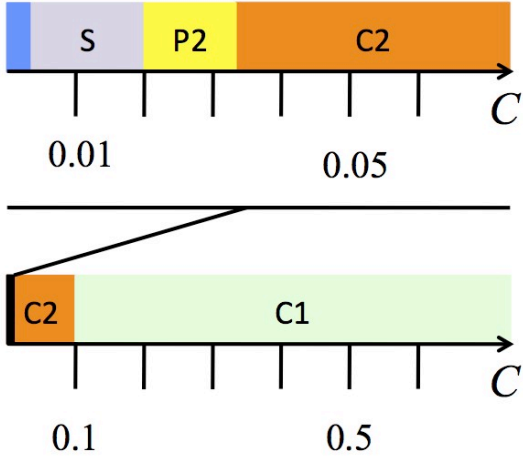

**Figure 2: Phase diagram of behavior of the network as a function of the coupling strength $C$. The blue region on the top line corresponds to regular unsynchronized oscillations, S to regular synchronized oscillations, P2 to period-two synchronized in anti-phase, C2 to chaos with period-two, and C1 to period-one chaos.**

For a very low coupling constant ($C = 0.001$), the units show regular oscillations that are desynchronized with the other units (shown in figure 3a), and for larger coupling strength the units again show regular oscillations that synchronize (figure 3b). We see that all units are synchronized with each other but there is phase lag to the regular oscillations of $x_b$. Here we use the term synchronization in the same sense as in synchronization by common noise of otherwise non-interacting oscillators [16, 17].

Then at $C = 0.02$ we see period doubling (figure 3c). The synchronized population divides into two with period doubling, that is a high and a low amplitude oscillation, the high amplitude oscillation of one subpopulation is synchronized with the low amplitude oscillation of the other subpopulation, and both show some lag to the oscillations of $x_b$. Then at $C = 0.031$ there is a further period doubling in which the subgroups split further in two different groups with alternating high or medium-high oscillations (figure 3d). For clarity, the behavior of a single oscillator is shown in figure 4.

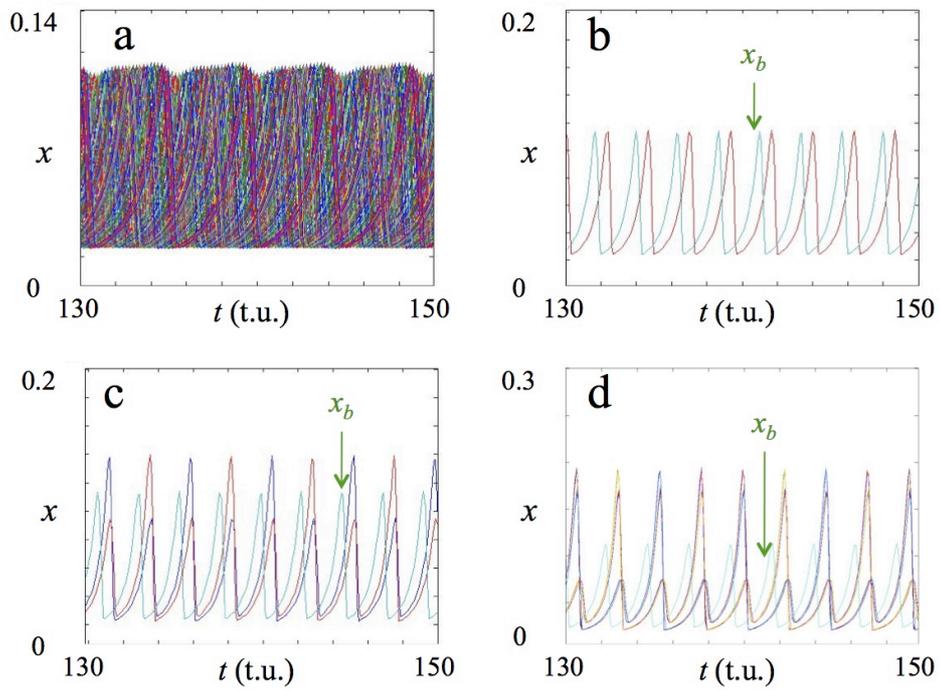

Figure 3: (a) Unsynchronized regular oscillations with $C = 0.001$, (b) synchronization of oscillators (blue and yellow) with a phase lag to $x_b$ with $C = 0.01$, (c) period doubling with two populations synchronized anti-phase to each other and with a phase lag to oscillations in $x_b$ with $C = 0.02$, (d) period-four $C = 0.031$.

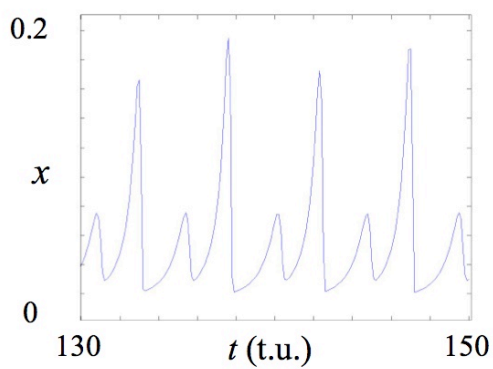

Figure 4: Behavior in time of a single oscillator at $C = 0.031$.

Then by increasing the coupling strength further, each oscillator behaves chaotically, however there is some kind of period-two oscillations for the whole system as seen in figure 5a. In figure 5b, the behavior of a single oscillator is compared to the oscillations of the common bath, where it is seen how the chaotic oscillator does not keep a constant phase difference to the oscillations of $x_b$.

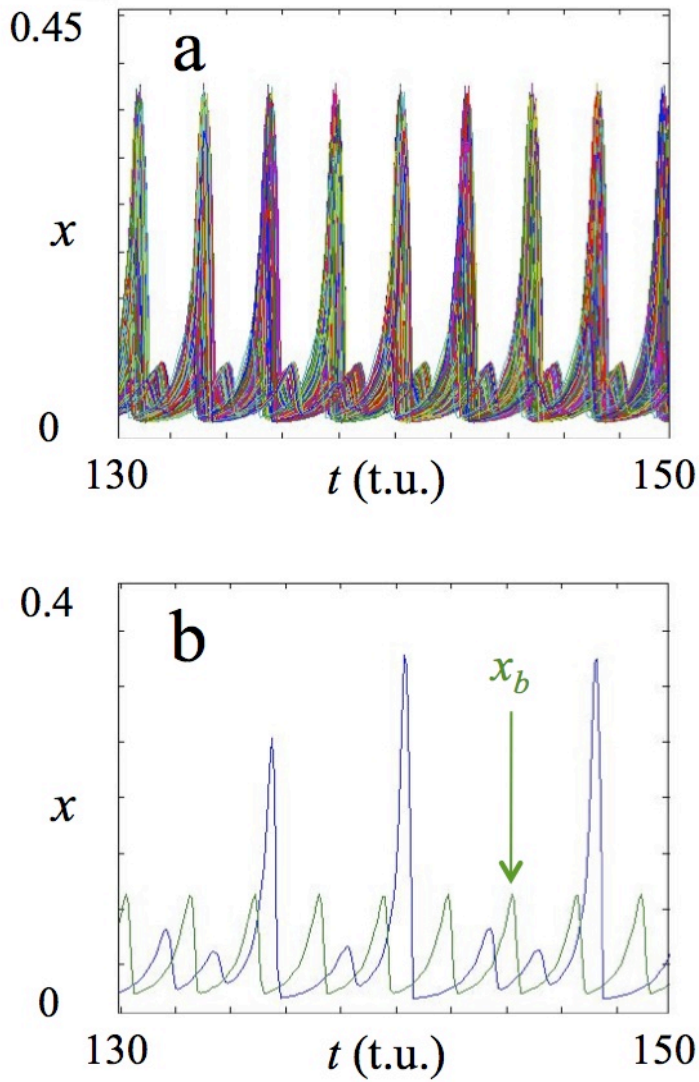

**Figure 5: Coherent behavior for the large population (a) of the sum of chaotic oscillators, one is shown in blue in (b) along the behavior of $x_b$ in green for $C$ = 0.1.**

For $C = 0.2$ or higher, the period-two oscillations of the global system turn into period-one oscillations, while each oscillator behaves chaotically. This is illustrated in figure 6.

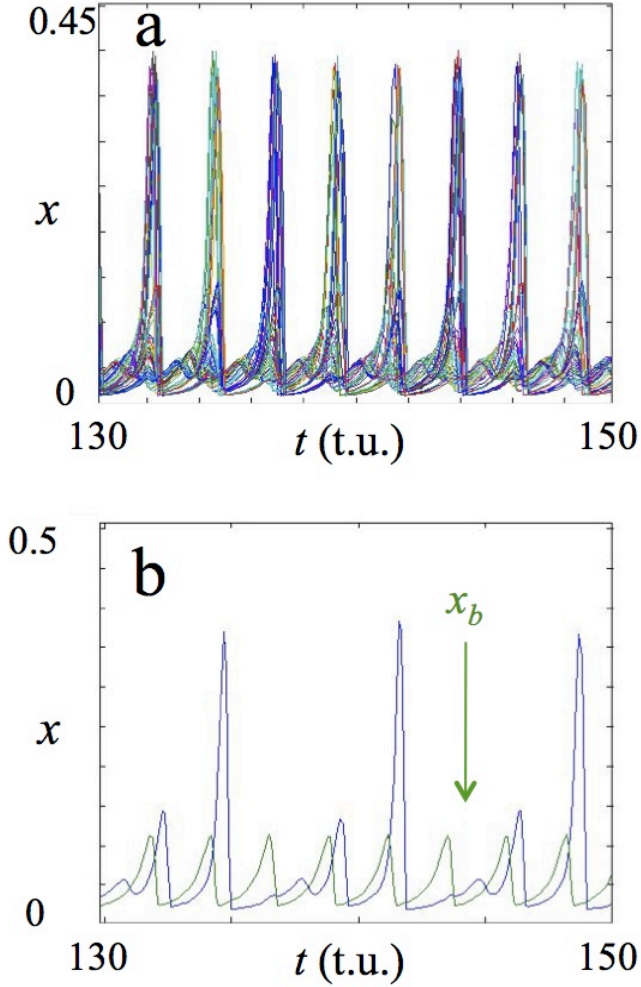

**Figure 6: Coherent behavior for the large population (a) of the sum of chaotic oscillators, one is shown in blue in (b) along the behavior of $x_b$ in green for $C = 0.9$.**

## IV. Discussion

It is perhaps counterintuitive that chaos emerges at a grater coupling strength that what is needed for synchronization of the system. We can understand the synchronization at low coupling strength by taking into consideration that the oscillations have constant amplitude, which corresponds to Kuramoto's approximation [18] and since the amplitude of oscillations is small in the regime of *defect-mediated turbulence* [9] only a small perturbation is needed. For comparison, we look at the synchronization in a large network where we make $f = 2$, keeping all the rest the same, so that it is not in the region of defect turbulence but where in reaction-diffusion the pattern corresponds to bulk oscillations [9]. The system needs a higher coupling strength to synchronize, and we see that at C = 0.1 the system is going towards synchronization but does not completely synchronize within the total time of the simulation (figure 7a). At higher coupling strength, the time to reach this complete synchronization diminishes as seen in figure 7b in comparison to figure 7a. Again here, the amplitude of the oscillation is constant. A system with small amplitude oscillations ($f = 2.1$) is than easier to influence that the system with high amplitude ($f = 2$).

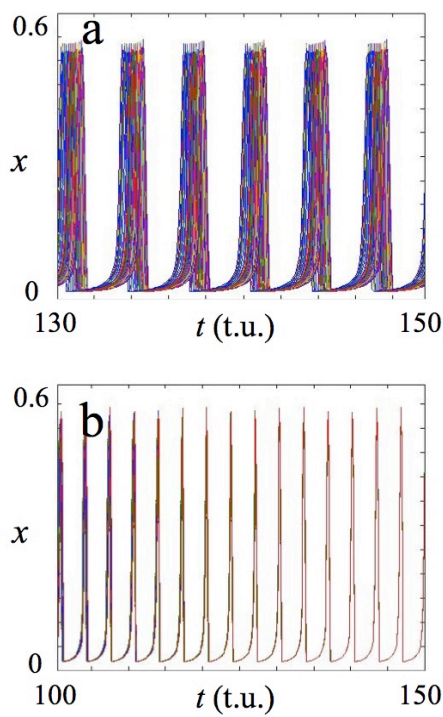

**Figure 7: Time behavior for all oscillators with C = 0.1 (a) and C = 0.5 (b). Other parameters $f = 2$, $q = 0.017$ and $\varepsilon = 0.04$.**

Period doubling is a well-known route to chaos, and here we see a further example. Perhaps more interesting is the amplitude only varying transition towards complete chaos seen in figure 3d. This state is somewhat similar to what is seen in chimera states, where part of the system is synchronized or coherent and part of it is chaotic [19]. There are different forms of chimeras and a classification scheme [20]. One of the forms is what is known as *amplitude chimeras*, where some population oscillates in phase with coherent amplitude while the other shows oscillations with incoherent amplitude [21]. and our system here shows this type of incoherent amplitude behavior. This route to chaos through change in amplitude contrasts what is seen in reaction-diffusion, where before *defect-mediated turbulence* the chaos takes the form of *phase chaos* [8]. In a sense, these results are also somewhat similar to synchronization of non-identical oscillators in which the phase shows synchronization but the amplitudes behave chaotically [22].

The discussion so far can be applied to a single oscillator coupled to a regular oscillator. However, it is interesting noting how the collective behavior of chaotic systems displays non-chaotic behavior. It is also interesting that at such low scale, namely a single oscillator under periodic forcing can generate chaos, which is in contrast to reaction-diffusion where the lifetime of chaos grows exponentially with the size of the system [23, 24]. A reaction-diffusion system with the parameters shown here, where only the activator is able to diffuse, shows some form of spatiotemporal chaos in which defect-mediated turbulence and traveling waves alternate in space and time (figure 8a). These traveling waves have higher amplitude than the defect state (figure 8b), and come from what it seems to be some kind of hard excitation. For small sizes, the system does not show chaotic behavior but rather traveling waves (not shown).

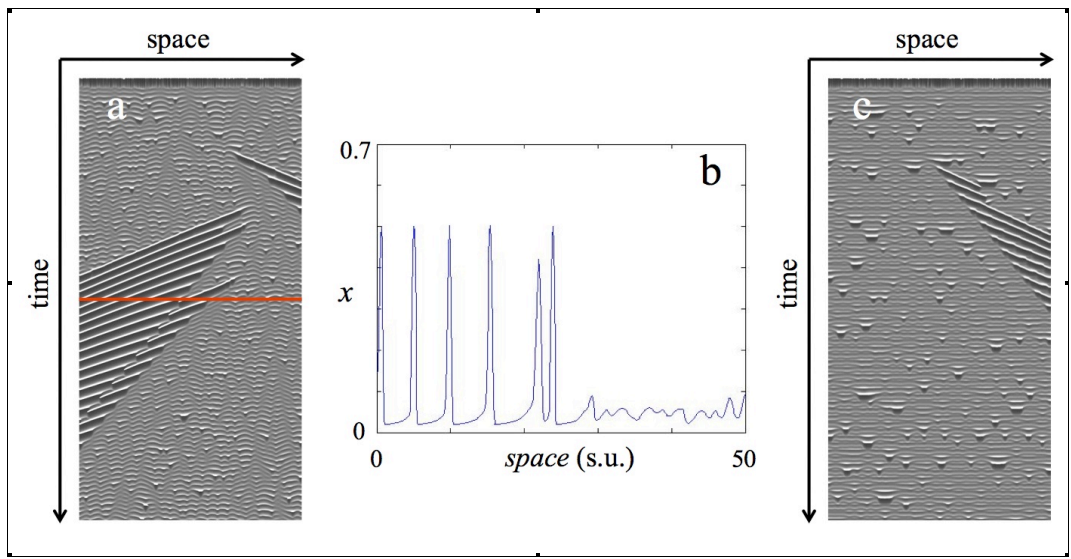

**Figure 8: (a) Space-time behavior, showing in white high concentration of $x$, with $f = 2.1$, $q = 0.017$ and $\varepsilon = 0.04$, $D_x = 0.1$ and $D_z = 0$. (b) Shows the concentration of $x$ along the red line in (a). Coupling strength is $C = 0.04$ in (a) and $C = 0.9$ in (c).**

On a single side fed reactor, the behavior of the feeding chamber can have and effect on patters, as seen in traveling spiral waves in the chlorine dioxide-iodine-malonic acid reaction where higher frequency of the feeding chamber produces breathing waves [25]. In the case at hand, coupling to the regular oscillation of the bath does not produce much of an effect, the appearance of traveling waves is reduced but not always completely inhibited, as exemplified in figure 8c, where the coupling strength is set to 0.9.

## V. Conclusions

Using the coupling strength through the activator as a parameter yields chaotic behavior beyond the coupling strength required for synchronization. This higher coupling induces changes in the amplitude of the oscillators, which, through a period doubling cascade brings the chaotic behavior.

Also, when only the activator diffuses within a reaction-diffusion domain, the behavior is different, namely the spatiotemporal chaos is the form of spatiotemporal intermittency between defect-mediated turbulence and traveling waves instead of only the defect-mediated turbulence seen when both activator and inhibitor diffuse equally [9].